\documentclass[showpacs,preprint]{revtex4}

\usepackage[latin1]{inputenc}
\usepackage{graphicx}

\newcommand{\tm}{\ensuremath{\mathcal{T}}}

\newcommand{\ket}[1]{\ensuremath{| #1 \rangle}}
\newcommand{\braket}[2]{\ensuremath{\langle #1| #2 \rangle}}
\newcommand{\gcpf}{\ensuremath{\mathcal{Z}}}
\newcommand{\Tr}{{\rm Tr\ }}
\begin{document}

\title{The statistical mechanics of the two-dimensional hydrogen-bonding self-avoiding walk including solvent effects}

\author{D P Foster}
\author{C Pinettes}

\affiliation{Laboratoire de Physique Th\'eorique et Mod\'elisation
(CNRS UMR 8089), Universit\'e de Cergy-Pontoise, 2 ave A. Chauvin
95302 Cergy-Pontoise cedex, France}

\begin{abstract}
A two-dimensional square-lattice model for the formation of secondary structures in proteins, the hydrogen-bonding model, is extended to include the effects of solvent quality. 
This is achieved by allowing configuration-dependent nearest-neighbour interactions. The phase diagram is presented, and found to have a much richer variety of phases than either the pure 
hydrogen-bonding self-avoiding walk  model or the standard $\Theta$-point model.
\end{abstract}

\pacs{05.40.Fb, 05.20.+q, 05.50.+a, 36.20.-r,64.60.-i}

\maketitle

\section{Introduction}

Self-avoiding walk models have been used for many years as models of real polymers in solution\cite{Gennes-P-G:1979sh,Vanderzande:1998ce,Cloiseaux:1990mi}. The 
thermodynamical behaviour of a linear polymer in  a good dilute solvent is dominated by its entropy,
which may be well modelled by an excluded volume interaction, leading to the idealised model of
a self-avoiding walk on a lattice. As the temperature of the polymer is lowered, typically the quality  of
the solvent is degraded, and the difference in affinity between the monomers (chemical building blocks of the polymer) and between the monomers and the solvent molecules becomes important. At low enough temperatures the polymer collapses and will precipitate from solution. This difference in affinities  
may be modelled in the self-avoiding walk model by an effective attractive interaction between neighbouring steps of the walk. The high temperature (good solvent) and low temperature (bad solvent) regimes are separated by a phase transition point known as the $\Theta$-point.

\begin{figure}
\begin{center}
\includegraphics[width=10cm,clip]{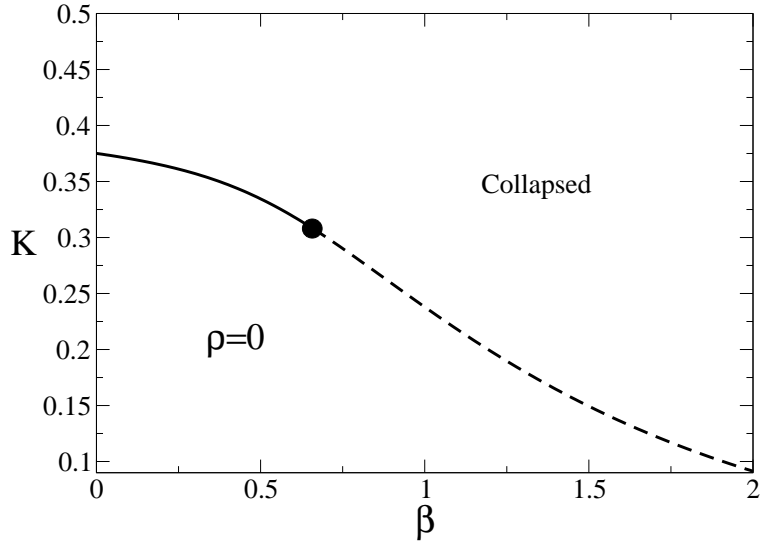}
\end{center}
\caption{The phase diagram for the standard $\Theta$-point model, showing a low-$K$
zero density (finite walk length) phase and a  high-$K$ (critical) collapsed phase, where the walk 
density is finite. At high temperatures (low $\beta$) the transition is second order (solid line),
 whilst at low temperatures (high $\beta$) the transition becomes first order (dashed line). These two behaviours are separated by a tricritical point, the $\Theta$ point.}\label{thetaschem}
\end{figure}

The canonical model for this system is the $\Theta$-point model, which consists of placing the interactions between nearest-neighbour lattice sites which have been visited non-consecutively by the self-avoiding walk\cite{Wall:1961tf,Domb:1974cq}. Typically this model is studied in the grand-canonical ensemble where the length of the walk is governed by a step fugacity $K$. The grand-canonical partition  function, \gcpf, is then written:
\begin{equation} 
\gcpf=\sum_{\rm walks} K^N\tau^{N_I},
\end{equation}
where $N$ is the length of the walk, $N_I$ are the number of nearest-neighbour interactions and $\tau=\exp\left(-\beta \varepsilon\right)$, $\beta=1/kT$, $\varepsilon<0$ is the (attractive) energy 
gained per nearest-neighbour contact. This model describes well the behaviour of simple linear polymers in solution. The phase diagram in the fugacity/temperature plane is shown in figure~\ref{thetaschem}. 
It is expected that any typical size of the walk, such as the radius of gyration, should scale as a power law with the length of the walk as follows: 
\begin{equation}
R_G\sim \langle N \rangle^\nu,
\end{equation}
where $\nu$ is simply the correlation length exponent defined in magnetic models.
Indeed, the self-avoiding walk model may be mapped onto an $O(n)$ symmetric spin model
in the limit $n\to 0$\cite{Gennes:1972xh}. It may then be seen that the phase transition line shown in figure~\ref{thetaschem}
is the line on which the average length diverges, for large temperatures (small $\beta$) continuously and for low temperatures discontinuously. The $\Theta$-point is then identified with a tri-critical point.

Whilst this model is of most practical importance in three dimensions, it has been extensively studied in
two dimensions, 
because it is expected that the critical behaviour is richer in two dimensions, particularly since the upper critical dimension of a tri-critical point is 3.

 An interesting question arises: the basic ingredients modelled by the $\Theta$-point model which 
 enable it to capture the essence of the real polymer in solution are the modelling of the entropic repulsion by the excluded volume interaction (self-avoidance) and the modelling of the short ranged attraction between the monomers. It should be expected, then, that any consistent way of modelling 
these two features would lead to a thermodynamically equivalent model. With this in mind, Blöte and Nienhuis\cite{Blote:1989rr} introduced a variant on the model in which the self-avoidance restriction is relaxed in that the
 walk can now visit lattice sites more than once, however the walk is not allowed to visit the lattice
 bonds more than once, and the walk is not allowed to cross itself. The interactions are now introduced
 for the doubly visited sites. Surprisingly this model has a phase diagram which differs from the 
 $\Theta$-point model in many important respects: there is an extra phase transition in the dense region of the phase diagram. This phase line is in the Ising universality class with a value of 
 the correlation length exponent $\nu=1$\cite{Blote:1989rr,Guo:1999rq}. The new collapse transition is not in the same universality
 class as the $\Theta$ point, having  an exponent $\nu=12/23$\cite{Warnaar:1992sr}, as compared to $\nu=4/7$. 
 At first sight these differences seem to be in contradiction with the idea of universality, which is required
 if we are even to think of modelling a polymer in solution by a lattice based walk model. In fact, 
 Universality is not violated. If the walk fills the lattice with a finite density, and its fractal (Haussdorf) dimension is the same as the lattice dimension, then the walk sees the lattice, and may be subject
 to lattice frustration effects. At the $\Theta$ point the density is zero, whilst it was shown that the
 density at the collapse transition in the Nienhuis-Blöte model is non-zero\cite{Foster:2003sh}. The presence of 
 an Ising transition in the dense region is an indication that the lattice interactions tend to pick out a 
 preferred direction, here corresponding to the lattice diagonals.
 
\begin{figure}
 \includegraphics[width=10cm]{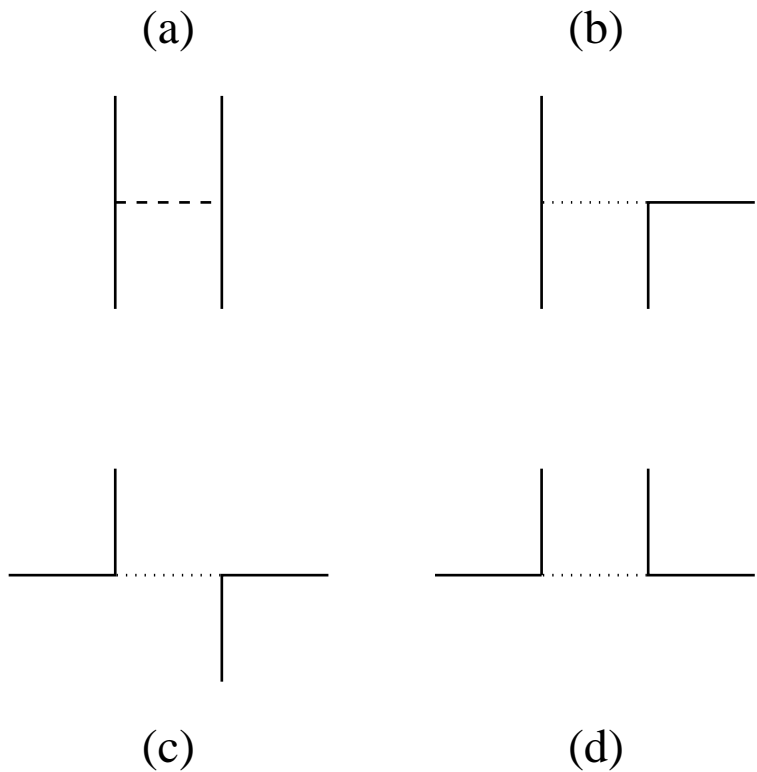}
 \caption{The nearest-neighbour interactions are split into two classes, those of type (a) where four bonds forming two parallel lines model the hydrogen bonds, whilst the others (b,c and d) model the solvent
 interactions. Configuration (a) induces a preferred orientation, whilst the other configurations do not.}\label{hint}
 \end{figure}

 Other lattice models have been introduced which contain collapse transitions, notably the bond-interacting self-avoiding walk\cite{Stilck:1996rq,Buzano:2002hc,Foster:2007fk} and the Hydrogen-bonding self-avoiding walk\cite{Bascle:1993fr,Foster:2001xd}. The first is simply the $\Theta$ point model in which the interactions are now between the nearest-neighbour visited lattice 
 bonds. The   Hydrogen-bonding self-avoiding walk was introduced to model the formation of secondary structures in proteins under the influence of the hydrogen-bond. Hydrogen bonds are induced by dipole-dipole interactions, and impose an orientation on the interacting portions of the polymer. How these interactions are implemented in the Hydrogen model is shown in figure~\ref{hint}. 
 
\begin{figure}
\begin{center}
\includegraphics[width=10cm,clip]{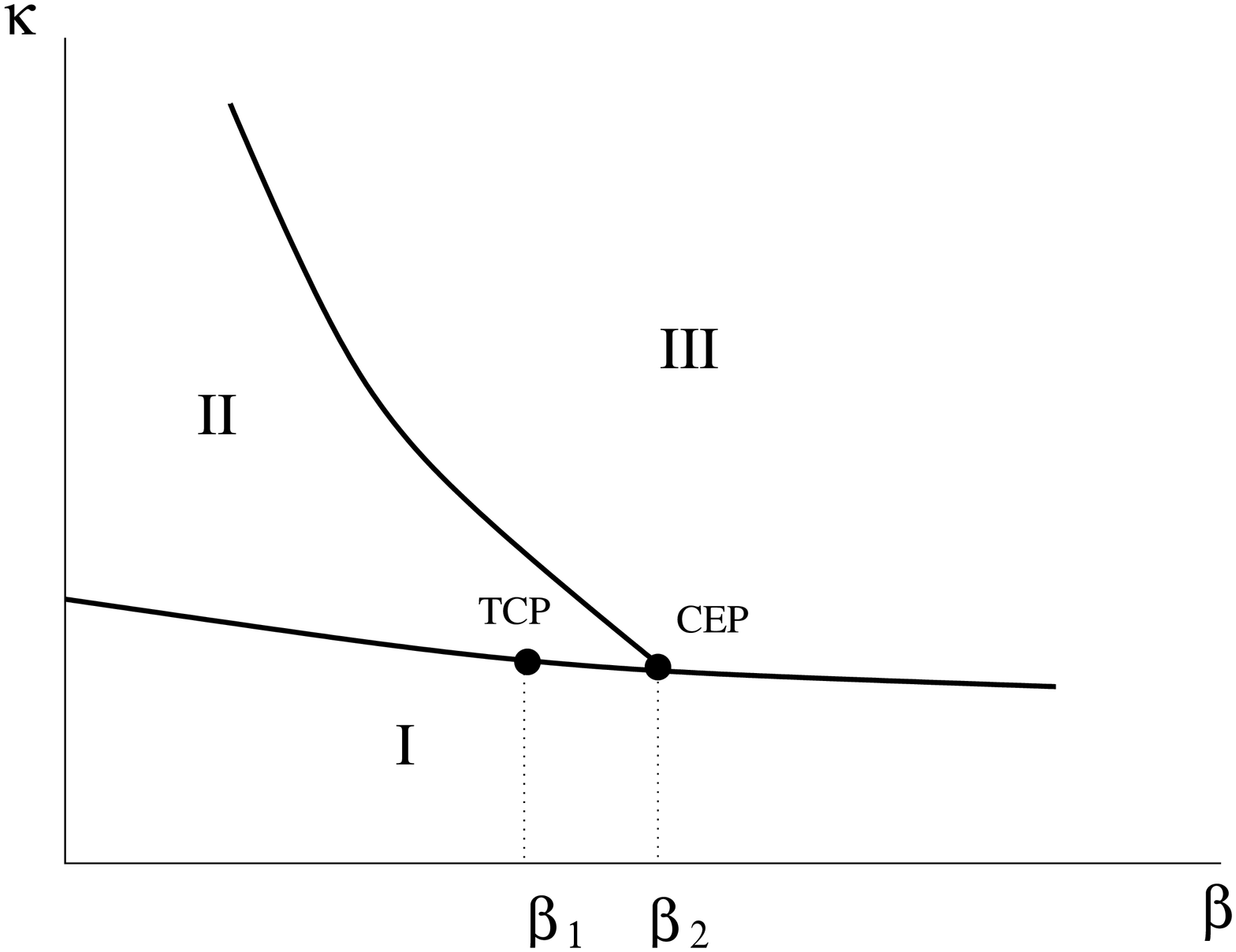}
\end{center}
\caption{A schematic version of the phase diagram 
for the bond-interacting self-avoiding walk model,
proposed  by  Buzano and Pretti\cite{Buzano:2002hc}
and confirmed by Foster\cite{Foster:2007fk} . Phase I is the low-K finite walk phase, II is the critical collapsed (liquid) phase and III is the crystalline oriented phase.}\label{biphase}
\end{figure}

The bond-interacting self-avoiding walk has been studied using mean-field type calculations
on the Bethe lattice\cite{Buzano:2002hc} and the Husimi lattice\cite{Stilck:1996rq}. 
These different studies have 
proposed radically different phase diagrams. A recent transfer
matrix calculation\cite{Foster:2007fk} indicates that the correct phase diagram is that proposed by Buzano and Pretti\cite{Buzano:2002hc}, and shown 
schematically in figure~\ref{biphase}. Whilst there is a collapse transition in the same universality
class as the standard $\Theta$ point, there is also the presence of an additional transition line 
in the dense region of the phase diagram. Unlike the Nienhuis-Blöte model, this transition line 
seems to be exotic in nature, with a non-divergent susceptibility\cite{Foster:2007fk}.  

The Hydrogen-bonding model was also investigated using transfer matrices\cite{Foster:2001xd}. The phase diagram 
was schematically similar to the Nienhuis-Blöte model, but the collapse transition was found to be first order. The high density transition line seemed not, in this study, to be in the Ising universality class, but the error bars were so large as to make it hard to rule out any possibility. We return to this question in this work, and using recently developed corner-transfer-matrix renormalisation-group (CTMRG) 
methods\cite{Foster:2003mb} we
manage to give an accurate estimate of the critical exponent $\nu$, clearly ruling out any possibility
that it could be in the Ising universality class.

Faced with this variety of different behaviours, it almost seems that any change to the model leads
to different behaviour for dense interacting self-avoiding walks. In order to investigate the relationship
between these different behaviours we introduce an extension of the Hydrogen-bonding self-avoiding 
walk model to include $\Theta$-type interactions. These interactions are shown in figure~\ref{hint}.
There is also a more direct motivation for these interactions; whilst a protein folds under the influence
of the hydrogen bonds, it is still subject to the quality of the solvent. Indeed it is the quality of the
solvent (or physiological temperature) which decides whether or not 
a protein is functional.  We will show that
this enlarged model contains several, if not all, of the different thermodynamic behaviours found above,
and provides a unifying framework in which to understand the effect of geometrical frustration in 
lattice walk models.

In the next section the model is presented. In section~\ref{results} results are first presented for two 
cases, showing two different behaviours. The results found then enable a mapping of the full
phase diagram. The article ends with a discussion of the results. 

\section{The model}\label{themodel}

The model studied in this article involves the self-avoiding walk on the square lattice with interactions between non-consecutive visited nearest-neighbour sites on the lattice. 
Unlike the standard $\Theta$-point model, the interactions are split into two sets, as shown in figure~\ref{hint}, between those which specify a particular direction, the hydrogen bonds, and those that do not, the solvent interactions. Hydrogen bonds carry an interaction energy $-\varepsilon_H$ 
and the others carry an interaction energy $-\varepsilon$. The thermodynamic behaviour may be investigated by introducing the grand-canonical partition function, \gcpf, from which many of the 
relevant thermodynamic quantities may be calculated. The grand-canonical partition function is given by:
\begin{equation}
\gcpf=\sum_{\rm walks} K^N\exp\left(\beta\left(N_I\varepsilon+N_H\varepsilon_H\right)\right),
\end{equation}
where $N_I$ are the number of solvent interactions, and $N_H$ are the number of hydrogen bonds. The fugacity, which controls the average length of the walk, is denoted by $K$, and $N$ is the total length of the walk. For convenience we define $\alpha=\varepsilon/\varepsilon_H$, and 
without changing the physics of the model, we may set $\varepsilon_H=1$; this simply sets the temperature scale.  The partition function then becomes:
\begin{equation}
\gcpf=\sum_{\rm walks} K^N\exp\left(\beta\left(N_H+N_I\alpha\right)\right).
\end{equation}

The primary tool we use in this article to map out the phase diagram of this model is the transfer matrix.
The transfer matrix method involves studying the model on a lattice strip of length $L_x \to \infty $ and width $L_y$, finite. 
In its simplest form, the model is considered on a lattice with periodic boundary conditions in both the $x$ and $y$ lattice directions. In this case the partition function may be rewritten in terms of a matrix product:
\begin{equation}
{\cal Z}=\Tr{\cal T}^{L_x},
\end{equation}
where $\mathcal{T}$ is the transfer matrix, which contains all the factors  required to take account of all
possible walk configurations between any two given lattice columns. Details on the transfer matrix
method used in this article may be found in reference~\onlinecite{Foster:2007fk}.

The partition function may then be expressed in terms of the eigenvalues $\lambda_i$, of the transfer matrix $\cal T$:
\begin{equation}
{\cal Z}=\sum_i \lambda_i^{L_x}.
\end{equation} 
The dimensionless free energy per spin is given by
\begin{equation}
f=\frac{1}{L_xL_y}\log\gcpf.
\end{equation}

In general the largest eigenvalue  is non-degenerate, and the sum is dominated by this largest eigenvalue, $\lambda_0$, giving, in the limit $L_x\to\infty$,
\begin{equation}
f=\frac{1}{L_y}\log\lambda_0.
\end{equation}
The problem is now reduced to studying the behaviour of the thermodynamic quantities
as a function of the width, notably using finite-size scaling techniques. In what follows we will drop the subscript $y$ and denote the lattice width by $L$. Once the free energy has been calculated, 
other quantities of interest can be calculated by taking suitable derivatives, for example the density of the 
walk on the lattice is given by 
\begin{equation}
\rho=\frac{\langle N \rangle}{L_xL_y}=K \frac{\partial f}{\partial K}.
\end{equation}
It is however possible to calculate such quantities directly from the eigenvalues and eigenvectors of the transfer matrix.
To see this, it is
necessary first to calculate the probability of having a given walk configuration
${\cal C}_x$ in column $x$. This probability is simply the ratio of
the partition function restricted to having configuration ${\cal C}_x$ in column $x$
and the unrestricted partition function, which in terms of transfer
matrices may be written:
\begin{equation}
p({\cal C}_x)=\lim_{L_x \to\infty}
\frac{\Tr \left\{T^x | {\cal C}_x \rangle\langle {\cal C}_x |T^{L_x-x}\right\}}{\Tr T^{L_x}}.
\end{equation}
Writing $|{\cal C}_x \rangle$ in terms of the eigenvectors, $\ket{i}$ of $\tm$ gives:
\begin{equation}
p({\cal C})=\lim_{L_x\to\infty}
\frac{\sum_i\lambda_i^{L_x}\braket{i} {{\cal C}}\braket{{\cal C}}{i}}{\sum_i\lambda_i^{L_x}}
\end{equation}
\begin{equation}
p({\cal C})=\braket{0}{{\cal C}}^2,
\end{equation}
where the eigenvectors are normalised. The subscript $x$ may be omitted by invoking translation invariance.
 The density, for example, is then found using
\begin{eqnarray}\nonumber
\rho&=&\sum_{\cal C} \frac{N({\cal C})}{L} p({\cal C}) \\
&=&\sum_{\cal C} \frac{N({\cal C})}{L}\langle 0 | {\cal C} \rangle^2,
\end{eqnarray}
where $N(\mathcal{C})$ is the number of occupied lattice bonds in configuration
 $\mathcal{C}$. The susceptibility can then be calculated either by taking a derivative of the density,
or by calculating directly $\langle N^2\rangle$ {\it for the column}, and hence the fluctuation.  
The two methods give slightly different results for a finite width strip, but agree in the thermodynamic limit. In the present article we choose to calculate the fluctuation directly.

It is straightforward to show that the correlation length, $\xi$, depends on the largest two eigenvalues through
\begin{equation}\label{logcor}
\xi=\frac{1}{\log\left(\frac{\lambda_0}{|\lambda_1|}\right)}.
\end{equation}
If the two eigenvalues become equal in modulus, the correlation length diverges, which is characteristic
of the long-range order found at a critical point. For an integer spin model (Ising, XY, Heisenberg etc.), 
the transfer matrix is positive (all elements strictly larger than zero) and Frobenius' theorem states
that the largest eigenvalue is non-degenerate for finite matrices. This implies that the correlation length
may only diverge in the thermodynamic limit $L_x,L_y\to \infty$. In our case, however, 
the transfer matrix is sparse,
and may be block diagonalised into an odd and an even sub-matrix. The odd sub-matrix is the 
transfer matrix for the walks which cross the lattice in the $x$-direction an odd number of times, whilst the even sub-matrix is the transfer matrix of walks which cross the lattice an even number of times. We include in the even sub-matrix the empty lattice configuration. There is no mathematical reason why the
largest eigenvalues of the different sub-matrices should not coincide, and indeed the lines where this is 
the case may correspond to transition lines in the phase diagram, since they correspond to lines 
where the correlation length diverges. It is important to note, however, that the high-density isotropic
phase is a critical phase, in which $\xi\to\infty$ everywhere in the infinite lattice system, and so the
condition that the two eigenvalues become degenerate is not a foolproof argument, and must be 
used with care.  A standard method for finding phase transition lines in a transfer matrix calculation is 
to use finite size scaling in the form of Nightingale's renormalisation group method\cite{Nightingale:1976fd}, which is based
on the scale invariance expected close to critical points for large enough lattice sizes. It shows that when
solutions exist for the finite-size renormalisation equation
\begin{equation}\label{nrgcond}
\frac{\xi_L}{L}=\frac{\xi_L^\prime}{L^\prime},
\end{equation}
then these lines are candidate critical transition lines, although, again, such solutions may exist in
the high density critical phase without corresponding to transition lines.

The use of transfer matrices in the determination of the phase diagram is convenient, since the 
partition functions are calculated exactly for infinite strips for any value of the parameters given. 
Since the partition function is known exactly, there are no convergence problems, and the full
phase diagram may be mapped with relatively little effort. The main problem is the size of the
matrices, which grow exponentially with the lattice width. This strongly limits the maximal width
which may be used, here to $L=9$. Added to the fact that the model contains strong odd/even parity
effects, the number of sizes available to more advanced finite-size scaling methods is too small
to be of much use. In order to be able to use finite-size scaling to calculate critical exponents for
various transition lines, we decided to use a recently introduced implementation of the 
corner-transfer-matrix renormalisation group (CTMRG) method appropriate for lattice walk models\cite{Foster:2003mb}.
This method, related to the better known density-matrix renormalisation group (DMRG)
method, enables the calculation of thermodynamic quantities for large lattice sizes, in particular the density of monomers, for lattices with the restriction
that $L_x=L_y$. The large lattice sizes are achieved by iteration from smaller lattice sizes. At each 
iteration the phase space is optimally pruned such that the calculation remains within the
constraints of the available computer resources and the error on the quantities of interest is minimised.
For further details on the implementation of this model, please see reference\cite{Foster:2003sh} and references
therein.  
  
\section{Results}\label{results}

In this section we present the results obtained from the transfer matrix calculations, supplemented when
necessary with results from the CTMRG method. We start by studying the small $\alpha$ regime,
where the model is expected to behave like the pure Hydrogen model ($\alpha=0$), 
studied by Foster and Seno\cite{Foster:2001xd} 
using transfer matrices, and in the Bethe approximation by 
Buzano and Pretti\cite{Buzano:2002hc}. At the other end of the scale, when $\alpha=1$ we recover the pure $\Theta$-point
model. Considering values of $\alpha$ close to $\alpha=1$ we find new behaviour, not present
in either the Hydrogen model or the $\Theta$-point model. The results found 
permit the phase diagram
to be plotted, and finally we will present results for this phase diagram.

\subsection{Results for $\alpha=0.2$}

Foster and Seno\cite{Foster:2001xd} 
 studied the model for $\alpha=0$ using transfer matrices. It was shown that the behaviour
 of this model was different from the standard $\Theta$-point model. The collapse transition was now
 first order, corresponding to a jump in the density as the low-$K$ transition line is followed. 
  The model also presents a high-$K$ transition line separating the usual isotropic dense (liquid) phase
 from an anisotropic (crystalline) phase.
They gave evidence that this transition was critical, in contradiction to extended mean-field type calculations
 performed by Buzano and Pretti\cite{Buzano:2002hc} on the same model, which
 predicts a first order transition. 
 
 In this section we show that for $\alpha=0.2$ we recover a similar behaviour, showing that the 
 crystalline phase and its associated phase transitions persist over a range of values of $\alpha$.
Transfer matrix calculations are combined with those of the CTMRG method to obtain more
accurate results for the critical behaviour of the liquid-crystalline transition, which we confirm to be
of second order.
  
\begin{figure}
\begin{center}
\includegraphics[width=10cm,clip]{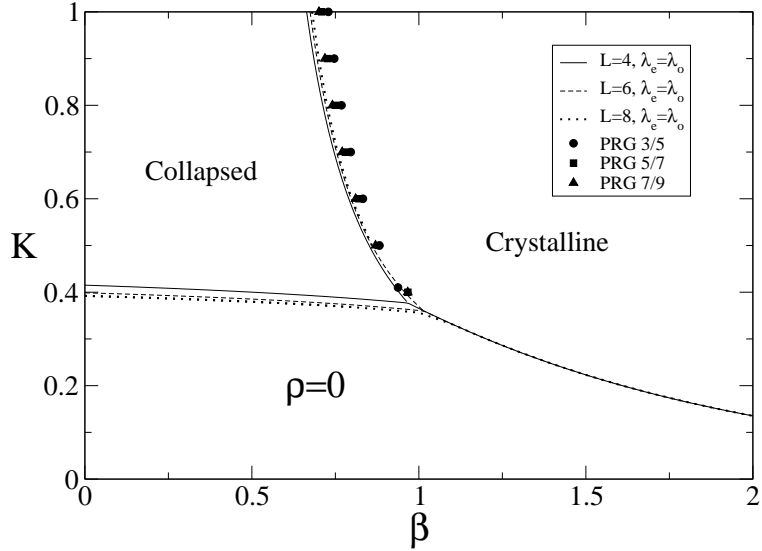}
\end{center}
\caption{Phase diagram for $\alpha=0.2$ calculated using eigenvalue crossings (lines), and the phenomenological RG method (points) for the high-$\rho$ transition.}\label{pda2}
\end{figure}

\begin{figure}
\begin{center}
\includegraphics[width=10cm,clip]{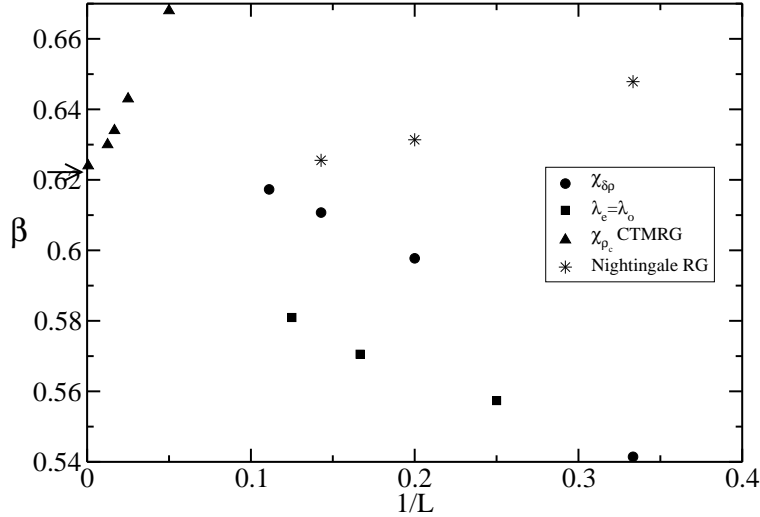}
\end{center}
\caption{Estimates for $\beta_c$ using various methods for $\alpha=0.2$ and $K=2$. Arrow shows the value found using
the data collapse method, $\beta_c=0.6222$. $\bullet$ represent the peaks of the fluctuations of $\delta\rho$, whilst the triangles correspond to the position of the peaks of the fluctuations in the corner density, calculated with CTMRG. $\star$ gives the position of the solutions to the Nightingale RG method, and the squares the position estimated using the condition that $\lambda_e=\lambda_o$.}\label{betaca2}
\end{figure}

 The phase diagram for $\alpha=0.2$ is shown in figure~\ref{pda2},  calculated using 
transfer matrices. In figure~\ref{pda2} the low-$K$ transition line between the finite-length $\rho=0$ 
phase and the dense phases is found by setting the largest eigenvalue of the transfer matrix to one. 
When even lattice sizes are considered, the largest eigenvalue corresponds to the largest eigenvalue from the odd sector of the
transfer matrix ($\lambda_o$) for small $\beta$ and the largest eigenvalue of the even sector ($\lambda_e$) when $\beta$ is large. The point where the
two cross, $\lambda_e=\lambda_o$, is identified as the crystallisation transition. This line extends both into the high and low
density phases. In the low density phase the line corresponds to the crossing of sub-dominant eigenvalues (since the largest eigenvalue, corresponding to the empty lattice, is $\lambda_0=1$).
This may be identified with a disorder line, indicating a change of local order. In the dense phase, however, the crossing corresponds to a crossing of the two largest eigenvalues, which from equation~(\ref{logcor}) may be seen to correspond to a divergent correlation length. This line may
then be identified with a special critical line where the long range order changes. The identification 
of this line with the phase boundary between the liquid and crystalline phases is not straightforward  
since the liquid phase is a critical phase, so it is possible that the line is buried within this phase. 
To verify that this is indeed the transition line, we compare the results with results calculated from 
Nightingale Renormalisation Group arguments (the points shown also in figure~\ref{pda2}). In figure~\ref{betaca2} we
give finite-size estimates for  the transition line for $\alpha=0.2, K=2$, showing the coherence
of the different methods, and verifying that the eigenvalues cross at the transition.

Transfer matrices are limited by the maximal lattice width that may be obtained, which in turn limits
the number of lattice sizes which may be used for studying finite-size behaviour. For this
reason we turn to the CTMRG method, which produces  results 
for large lattice sizes, permitting  better estimates
of the critical exponent, critical density and temperature. 
Unlike the transfer matrix method, where we dealt with infinite strips of finite width, 
in what follows we will be looking at a lattice finite in both directions, with $L_x=L_y=L$.

In order to use finite-size scaling, we consider  the scaling form for the density. 
Here it was convenient to fix $K$ and vary $\beta$, for which we expect the following scaling form:
\begin{equation}\label{scaling}
\rho_L(\beta)=\rho_\infty(\beta)+L^{1/\nu-2}\tilde{\rho}(|\beta-\beta_c|L^{1/\nu}).
\end{equation}

\begin{figure}
\begin{center}
\includegraphics[width=10cm,clip]{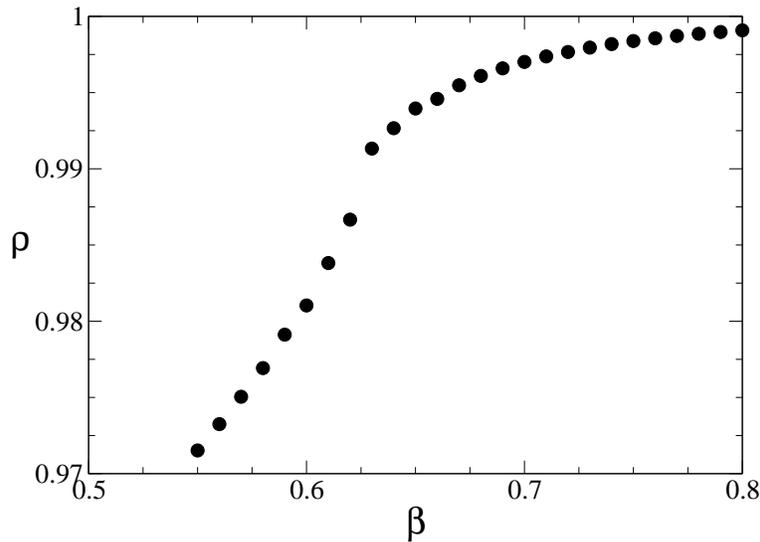}
\end{center}
\caption{Density calculated for $\alpha=0.2, K=2$ using CTMRG with $L=1000$ }\label{densa.2}
\end{figure}

\begin{figure}
\begin{center}
\includegraphics[width=10cm,clip]{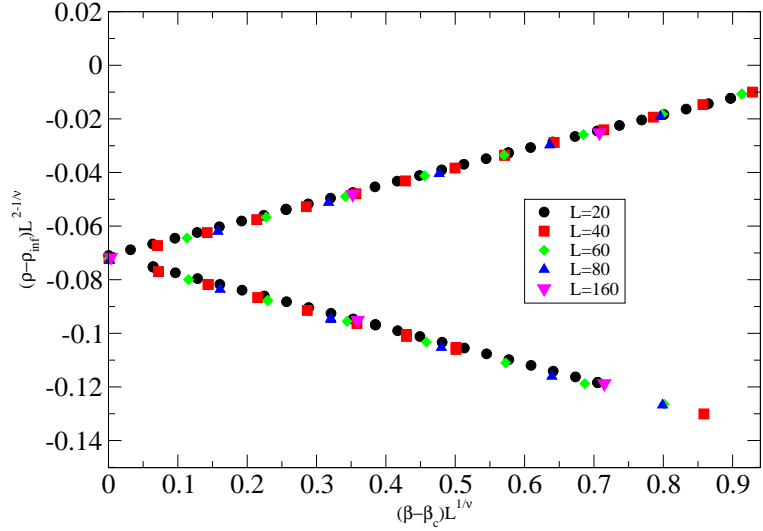}
\end{center}
\caption{Data collapse for the density close to the transition using data from the CTMRG method with  $\alpha=0.2, K=2$. The finite-size scaling form of the density is taken 
with $\rho_\infty=0.989$, $\beta_c=0.6222$ and $\nu=0.87$}\label{dca2}
\end{figure}

The scaling form given in equation~\ref{scaling} gives the leading behaviour close to the critical
point. The corrections to scaling may be expected to be negligible ``close enough'' to the critical
point. It is clear that if the following variables are plotted :
\begin{eqnarray}
x&=&|\beta-\beta_c|L^{1/\nu},\\
y&=&(\rho_L(\beta)-\rho_\infty)L^{2-1/\nu},
\end{eqnarray}
then, close enough to the critical point, the points plotted should fall onto the universal curve
$y=\tilde{\rho}(x)$. This phenomenon is known as data collapse. The values of $\beta_c$, $\nu$ and $\rho_\infty$ are not known, but can be determined by choosing values which give the best 
data collapse. It is of course useful to know some estimate of $\beta_c$ and $\rho_\infty$ by some other
method, estimates which we are able to improve by optimising the collapse of data close to the
transition. In figure~\ref{densa.2} we plot the density as a function of $\beta$ for $K=2$ and 
a lattice $L=1000$. It is readily seen that $\rho_\infty\approx 0.99$ and $\beta_c\approx 0.62-0.63$.
Starting with these initial values we determined values of the parameters which gave the best
data collapse, and we find: $\rho_\infty=0.989\pm 0.001$, $\beta_c=0.6222\pm0.0005$ and $\nu=0.87\pm0.02$. The resulting curve $y=\tilde{\rho}(x)$ is shown in figure~\ref{dca2}. The error bars correspond to the range of values over which the parameters may be varied before we clearly no longer
have collapse of the data. We limited the lattice sizes to $L\leq 160$ in the study, since, due to the
factor $L^{1/\nu}$ in the variable $x$, the points which appear in the figure are closer to the critical
temperature as the lattice size increases. The error in the determination of the point is also amplified
by the factor $L^{2-1/\nu}$ in $y$. These considerations limit the maximum size considered. 

\begin{figure}
\begin{center}
\includegraphics[width=10cm,clip]{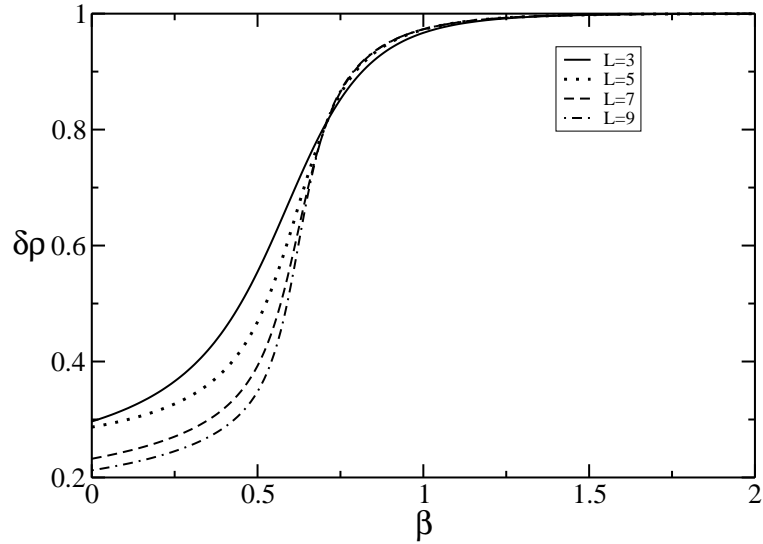}
\end{center}
\caption{Plot of $\delta\rho=|\rho_h-\rho_v|$ for $\alpha=0.2$ and $K=2$.}\label{dra2}
\end{figure}

Clearly, in the crystalline phase, the walk will wish to align with one of the lattice directions, with a tendency to eject corners from the bulk. Whilst we expect the density of corners to differ in the two phases, we note that the data collapse indicates that the density of the walk has not reached one, and so
the density of corners must be non-zero. This means that whilst the fluctuations in the numbers of 
corners should diverge at the transition, the corner density is not a good order parameter. On the other
hand the  
crystalline phase  is anisotropic, whilst the two other phases are isotropic. A natural order 
parameter is then $\delta \rho=|\rho_v-\rho_h|$, the difference between the densities of vertical and
horizontal bonds. In the isotropic phases this will vanish, but not in the anisotropic phase. We are not able to calculate this quantity with our CTMRG calculation, since the symmetries of the lattice are
explicitly used in the method\cite{Foster:2003mb}, but we have 
direct access to this parameter using transfer matrices, through
\begin{equation}
\delta\rho=\frac{1}{L}\sum_{\cal C}|N_v({\cal C})-N_h({\cal C})|p({\cal C}).
\end{equation}
This is shown in figure~\ref{dra2} for $\alpha=0.2, K=2$.
The peaks of the fluctuations in $\delta\rho$ and the density of corners, $\rho_c$, may be used as 
estimators for the liquid-crystal phase transition line. These are shown, along with other estimates, in figure~\ref{betaca2}.

\subsection{Results for $\alpha=0.8$}

\begin{figure}
\begin{center}
\includegraphics[width=10cm,clip]{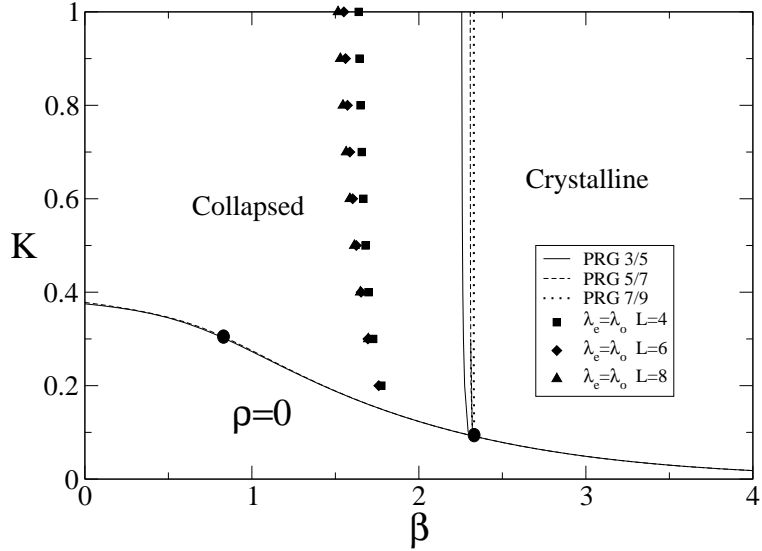}
\end{center}
\caption{Phase diagram for $\alpha=0.8$ calculated using Nightingale's Phenomenological RG method. The special transitions along the low-$K$ line are shown. The first is the $\theta$ point for $\alpha=0.8$ whilst the second is the collapsed-crystalline transition. The solution of the condition $\lambda_e=\lambda_o$ is shown for various sizes using points, and can be seen to give a distinct line,
which does not converge to the collapsed/crystalline transition line. }\label{pda8}
\end{figure}

In this section we choose to study the phase diagram for $\alpha=0.8$, where the model is found
to have a very different behaviour. The phase diagram calculated using the phenomenological renormalisation group is shown in figure~\ref{pda8}. Whilst we still find the three phases: the low-$K$ 
zero density phase, and the liquid and crystalline phases at higher $K$, the diagram is quite
different in appearance. 

\begin{figure}
\begin{center}
\includegraphics[width=10cm,clip]{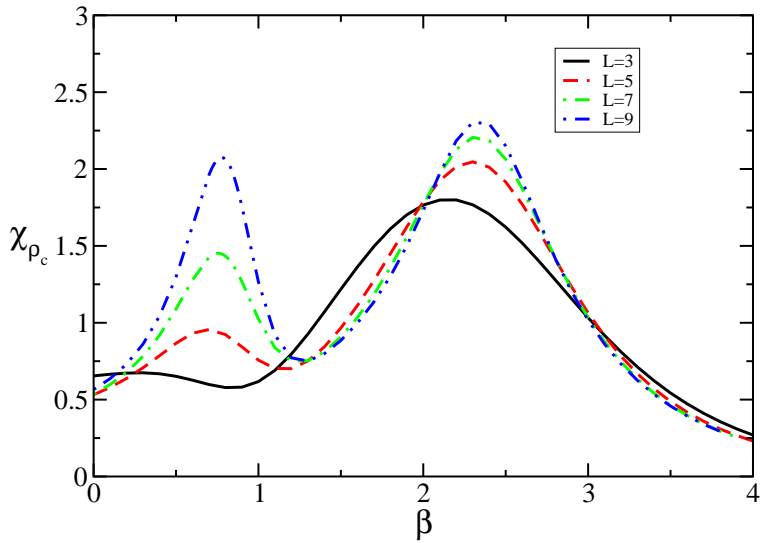}
\end{center}
\caption{Fluctuations of the number of corners in the walk for $\alpha=0.8$ calculated 
along the lower critical line showing clearly the existence of two transitions. 
Both transitions appear to be critical.}\label{flcta8}
\end{figure}

As the low-$K$ transition line is followed, we have first a $\Theta$ type transition, followed later by 
a crystallisation transition. The transition from finite walk to the dense phases is a second order
transition in the self-avoiding walk class for $\beta<\beta_{\Theta}$ becoming first order for $\beta>
\beta_{\Theta}$. In figure~\ref{flcta8}, the corner density fluctuations are plotted along the low-$K$ transition 
line. The corner susceptibility is calculated by introducing an additional parameter corresponding to 
a bending energy, and then applying the fluctuation dissipation  theorem. Defining $\beta_{\rm corn}=\varepsilon_{\rm corn}/kT$, we define
\begin{equation}
\chi_c=\frac{\partial \rho_c}{\partial \beta_{\rm corn}}.
\end{equation}

The formation of two peaks may clearly be seen, corresponding to the two special points along the low-$K$ line. Interestingly, unlike the $\alpha=0.2$ case, the two transitions appear to be 
critical. 

\begin{figure}
\begin{center}
\includegraphics[width=10cm,clip]{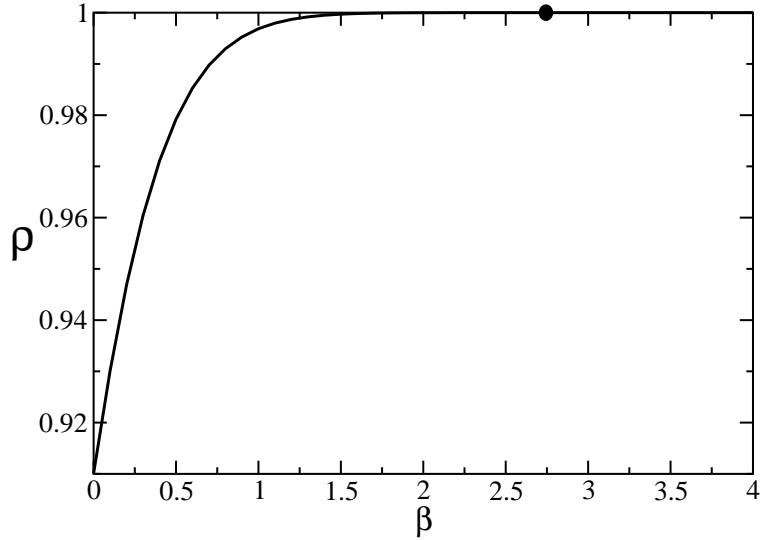}
\end{center}
\caption{Bond density calculated for $K=2$, $\alpha=0.8$ using CTMRG with $L=1000$. The circle indicates the location of the collapsed/crystalline phase transition calculated using data collapse (see figure~\ref{DCa8}). The transition can be seen to occur at a density $\rho=1$.}\label{densa8}
\end{figure}

\begin{figure}
\begin{center}
\includegraphics[width=10cm,clip]{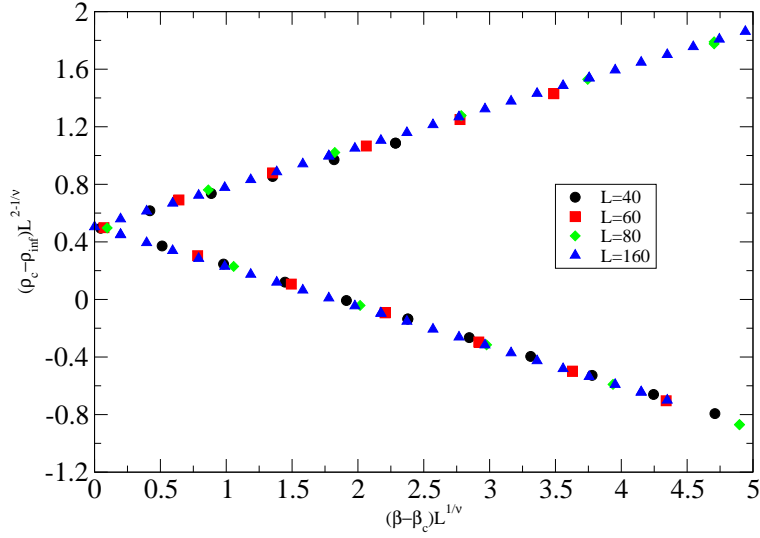}
\end{center}
\caption{Data Collapse of the corner density, $\rho_c$, for $\alpha=0.8$, $K=2$ fitted with $\rho_{c,\infty}=0.2744, \beta_c=2.349$, and $\nu=0.96$.}\label{DCa8}
\end{figure}

We now investigate the nature of the transition line between the two dense phases. What is interesting
is that the condition $\lambda_e=\lambda_o$ for even transfer matrices, 
which coincided with this line for $\alpha=0.2$, is well 
within the isotropic collapsed phase here. This is not contradictory, since the collapsed phase is
critical, and so the condition $\lambda_e=\lambda_o$ must correspond to a change of order within the
critical phase, but not to the phase boundary. 
This is the first indication that the transition line here is
different from the transition in the previous section. 
Here again we  turn to CTMRG and look for
the conditions for data collapse. 
The density, $\rho$, appears to have saturated to one (see figure~\ref{densa8}),  
making it impractical to use, however it is expected that the density of corners, $\rho_c$, should
scale in the same way, and this is what we use here.
The best fit was obtained for $\rho_{c,\infty}=0.2744\pm 0.0005, \beta_c=2.349\pm 0.003$, 
and $\nu=0.96\pm 0.02$, and is shown in figure~\ref{DCa8}.

\begin{figure}
\begin{center}
\includegraphics[width=10cm,clip]{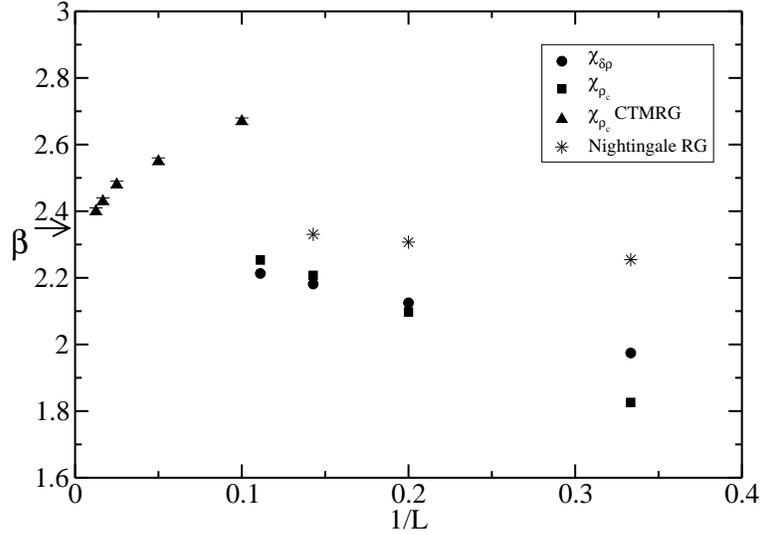}
\end{center}
\caption{Estimates for $\beta_c$ using various methods for $\alpha=0.8, K=2$. Arrow shows the value, $\beta_c=2.349$, found using
the data collapse method.}\label{betaca8}
\end{figure}

In figure~\ref{betaca8} we show different estimates for the critical point for $K=2$ and $\alpha=0.8$.

\subsection{The $\alpha-\beta$ phase diagram}

The full phase diagram is expressed in three variables, $K,\ \alpha$ and $\beta$, and is difficult 
to picture. In this section we present the phase diagram in the $\alpha-\beta$ plane 
calculated on the surface $K=K^*$, where 
$K^*$ is the value of $K$ required for the average length of the walk to just diverge. This is what is
generally calculated in Monte-Carlo simulations, and corresponds to the 
``long polymer in dilute solution" limit. 

\begin{figure}
\begin{center}
\includegraphics[width=10cm,clip]{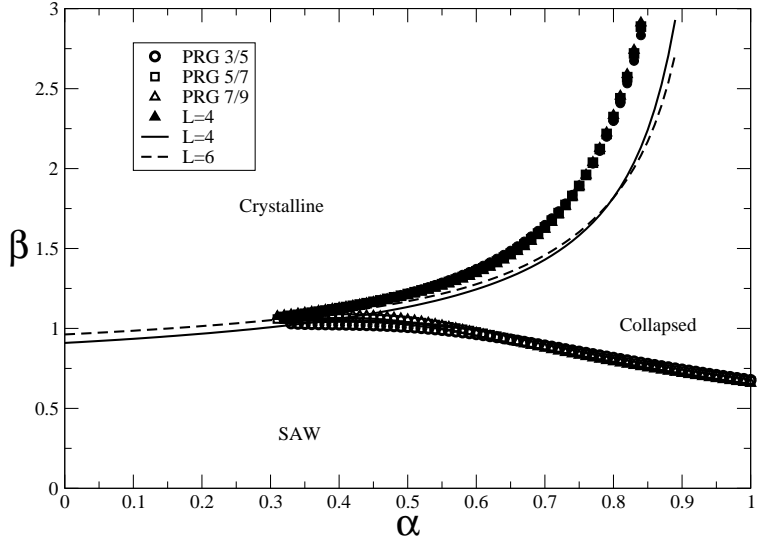}
\end{center}
\caption{Phase diagram in the $\alpha-\beta$ plane
calculated using phenomenological RG and eigenvalue crossings. The solid lines show the solutions to the condition $\lambda_e=\lambda_o$, expected to coincide with the self-avoiding walk--crystalline transition.}\label{phase1}
\end{figure}

When $\alpha=1$ the model corresponds to the pure $\Theta$-point model, with only one transition point on the $K=K^*(\beta)$ line; the $\Theta$ tricritical point. This point is easily found by looking at
the crossings of the finite-size estimates of $\nu$ calculated along the line of solutions to equation~\ref{logcor}. For small $\beta$ these estimates tend to $\nu_{SAW}=3/4$ whilst for large $\beta$ they tend to $\nu=1/2$, characteristic of the first order collapsed-walk line in two dimensions. 
In between these two behaviours we find a point, which tends to $\nu_{\Theta}=4/7$. By the way the estimates tend to their limiting values, this intermediate point shows up as a crossing in the different finite-size estimates. Looking at these estimates as a function of $\alpha$ leads to the extended line of tricritical points in the $\Theta$-point universality class.  This is the usual method for determining
the location of the tricritical point, but it requires the use of three lattice widths to determine one
estimate. Here we propose a different method. The low-$K$ transition line is determined by 
looking for solutions of the phenomenological RG equation (\ref{nrgcond}) with the correlation length
defined by equation (\ref{logcor}) taking $\lambda_0=1$ and $\lambda_1$ is the largest eigenvalue
taken from the odd or even sectors of the transfer matrix. A tricritical point has an additional 
correlation length which diverges, corresponding to the two relevant directions  in the renormalisation group sense. We look for the solutions of equation (\ref{nrgcond}) with a correlation length 
calculated using the largest eigenvalue from the odd and even sectors. This method only requires
two lattice widths to estimate the location of the tricritical point. Additionally, the high-$K$ transition line
is found using (\ref{nrgcond}) with these same two eigenvalues. The method described therefore also
locates the position of the crystallisation transition along the low-$K$ transition line. The phase
diagram calculated by this method is shown in figure~\ref{phase1}.  For small $\alpha$ there is no
solution. However this region of the phase diagram corresponds to the region which is expected
to behave like the pure hydrogen-bonding model, and therefore 
the condition $\lambda_e=\lambda_o$
corresponds to the first order transition line.

\begin{figure}
\begin{center}
\includegraphics[width=10cm,clip]{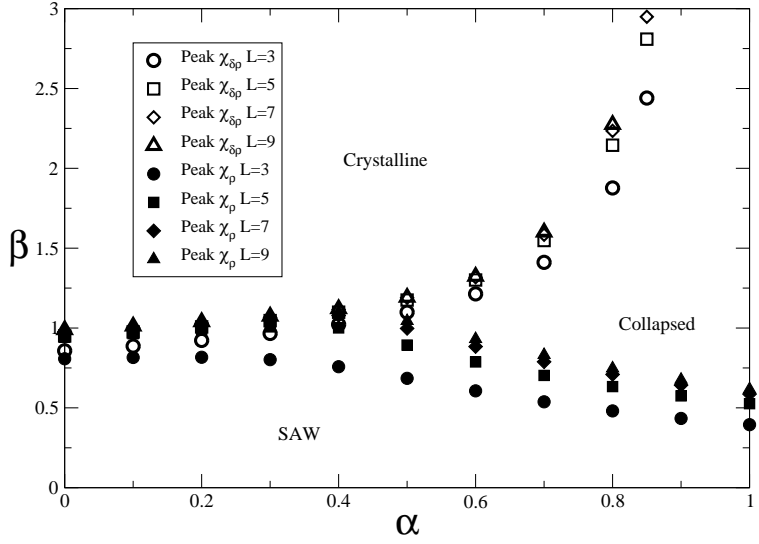}
\end{center}
\caption{Phase diagram in the $\alpha-\beta$ plane
calculated using the peaks of $\chi_\rho$ and $\chi_{\delta\rho}$ calculated using transfer matrices. The peaks of $\rho$ fluctuations pick out the line of $\Theta$ points, whilst the peak of $\delta\rho$ fluctuations pick out the transition between the isotropic collapsed phase and the anisotropic crystalline phase. These two lines merge to form the first order transition line separating the SAW phase from the crystalline phase.}\label{phase2}
\end{figure}

In figure~\ref{phase2} the phase diagram is calculated using the peaks of the two susceptibilities. The
upper line is calculated by looking for the peak of the fluctuations of $\delta\rho$ whilst the lower line
is calculated using the peaks of the fluctuations of $\rho$. For $\alpha$ less than about $0.4$ 
the two sets of lines merge and give estimates for the single hydrogen-bonding 
like first order line, whilst for values
of $\alpha$ larger than about $0.5$ the two sets of lines are distinct, the lower line corresponding to the
line of $\Theta$ like tricritical points, whilst the upper line corresponds to the critical crystallisation line.
Somewhere in the region $\alpha=0.4\to 0.5$ 
these two lines merge into a higher order multi-critical
point. 

Looking closely at figure~\ref{phase1} it may be seen that the upper and lower transition lines tend
to come together in the region
$\alpha=0.3\to 0.5$, with the lower line developing a plateau. It is probable that the multicritical point is not located at the cusp where solutions
end, but at a higher value of $\alpha$, probably in the same range of values. The methods employed
in this article were not able to determine this point more accurately. The CTMRG, which enables
larger sizes to be obtained, becomes impractical in this region, particularly when the number
of constraints required to define the point is considered. 

\section{Discussion}

There are now a number of similar models which display an 
anisotropic crystalline phase, with a variety of 
different types of high-density transition. 
The first is the vertex-interacting self-avoiding walk due to Blöte and Nienhuis\cite{Blote:1989rr}, which
displays an Ising like high-density transition with $\nu=1$. 
The Hydrogen-bonding model was shown to also have
a critical transition\cite{Foster:2001xd}, but in a different universality class, which is confirmed in this article, where the
value $\nu\approx 0.87$ is found. 
Lastly the bond-interacting $\Theta$-point model\cite{Foster:2007fk} which is conjectured to have a softer, 
higher order critical transition. 

What is interesting in the model presented here is that two different high-$K$ critical behaviours
are displayed in one model. For smaller values of $\alpha$ the transition from the collapsed to crystalline phase is of the Hydrogen-Bonding class, and occurs at densities which are close to $\rho=1$ but on close inspection we  clearly have $\rho<1$, as may be seen in figure~\ref{densa.2}.

Data collapse for $\alpha=0.8$ gave $\nu=0.96$ for the best fit, however the fitting was less clear than 
for $\alpha=0.2$ and it is possible that the correction terms are more important. The value of $\nu$ calculated would then be an effective exponent. It is tempting to conjecture that the true value of 
$\nu=1$, in analogy with the vertex-interacting model. However, as may be seen in figure~\ref{densa8}
the transition may be seen to occur well after the density saturates to $\rho=1$. If this is the case, then
the walk is essentially a Hamiltonian walk at the transition, and looks very much like the model with
a penalty for the formation of corners in the Hamiltonian walk limit studied by Saleur\cite{Saleur:1986on}, where 
he conjectured
that the transition was of infinite order BKT transition of the same type as in the F-model, but is in contradiction to the results
presented here. This contradiction was first seen in the Hydrogen model in the Hamiltonian limit\cite{Foster:2001xd}, where the transfer matrix results also gave estimates close to $\nu=1$. 
This is a point which warrants further investigation.

The results found here are to some extent confirmed by 
a Monte-Carlo study, mainly in three dimensions, which has 
appeared during the final stages of this work\cite{J-Krawczyk:fk}. Notably, based on a flat-PERM study,
a similar phase transition to that presented in figures~\ref{phase1} and~\ref{phase2} is found. 
The transition  from the collapsed phase to the crystalline phase along the surface where the walk length just diverges was seen to be probably critical, as is the case here.


\end{document}